\documentclass[10pt]{article}

\usepackage{amsmath}
\usepackage{graphicx}

\newcommand{\water}{H$_2$O}
\newcommand{\SR}{S\!R}
\newcommand{\Sr}{Sr}
\newcommand{\PSR}{P_{S\!R}}
\newcommand{\PSr}{P_{Sr}}

\newcommand{\dd}{\text{d}}

\begin{document}

\title{Orbits in the \water\ molecule}
\author{K. Efstathiou and G. Contopoulos\\
Center for Astronomy, Academy of Athens\\
Anagnostopoulou 14,
  GR-10673, Athens, Greece}
\maketitle

\begin{abstract}
  We study the forms of the orbits in a symmetric configuration of a
  realistic model of the \water\ molecule with particular emphasis on
  the periodic orbits. We use an appropriate Poincar\'e surface of
  section (PSS) and study the distribution of the orbits on this PSS
  for various energies. We find both ordered and chaotic orbits. The
  proportion of ordered orbits is almost 100\% for small energies, but
  decreases abruptly beyond a critical energy. When the energy exceeds
  the escape energy there are still non-escaping orbits around stable
  periodic orbits. We study in detail the forms of the various
  periodic orbits, and their connections, by providing appropriate
  stability and bifurcation diagrams.
\end{abstract}

\footnotetext[1]{Accepted for publication in \emph{Chaos}
 \textcopyright\ American Institute of Physics}

\vfill\eject

{\bf Small molecules are quantum systems, but their classical study
reveals their most important features. The method of the Poincar\'e
surface of section plays a prominent role in such a study.  A
difficulty that appears often in realistic systems is that one can not
always choose a flat Poincar\'e surface of section that intersects all
periodic orbits. In this paper we use a curved Poincar\'e surface of
section in order to study a realistic model of the \water\
molecule. We find the phase plots for different values of the energy
of the molecule. Then we find all the periodic orbits of period 1 and
2 and compute their stability and bifurcation diagrams.  }

\section{Introduction}

Two dimensional Hamiltonian systems are often studied using the
technique of the Poincar\'e map, where one reduces the four
dimensional, continuous time flow of the system to an associated
\emph{two dimensional} discrete map, by choosing an appropriate
surface of section. Usually a flat surface of section is chosen and
the orbits that cross this surface with a particular direction are the
consequents of the Poincar\'e map.

For many systems a flat PSS is not suitable since there are orbits of
these systems that do not cross the surface of section. In these cases
another choice must be made. In this paper we study a symmetric model
of the \water\ molecule where we have chosen a curved PSS. This choice 
for the PSS permits us to study all the periodic orbits of the system
for all the values of energy.

The study of periodic orbits and phase plots in realistic models of
simple molecules, such as the model we are studying in this paper,
gives interesting information that can be compared with the behavior
of the corresponding quantum system and with experiments
\cite{farantos90,keshavamurthy97}. In particular the
relation between classical periodic orbits and quantum mechanical
eigenfunctions was emphasized by several authors, starting with the
work of Gutzwiller \cite{gutzwiller71}. The most surprising result was
that in many cases one simple periodic orbit determines the basic form
of the eigenfunction
\cite{heller80,davis81,farantos85,founargiotakis89}.  Further
details can be found by using more periodic orbits \cite{berry89} and
the asymptotic structures around unstable periodic orbits
\cite{schweizer98}.

A partial study of periodic orbits in the \water\ molecule has been
done by Lawton and Child\cite{lawton81,lawton79}, Jaff\'e and
Brumer\cite{jaffe80} and by Kellman\cite{kellman95}, who emphasize the
bifurcation of the local stretching modes from the normal stretching
vibrational mode. These papers contain references to previous work on
orbits in the \water\ molecule.

We describe the model of the \water\ molecule in section 2, and use a
convenient set of coordinates in defining the Hamiltonian. We
calculate the equipotential surfaces for the symmetric molecule
(section 3) and in section 4 we define an appropriate Poincar\'e
surface of section and discuss its properties.

Then we study in detail the phase plots on such a Poincar\'e surface
of section and find the main islands of stability and the chaotic
zones (section 5). We calculate the main periodic orbits and their
bifurcations and stability in section 6. Finally we summarize our
conclusions in section 7.

\section{A model for the \water\ molecule}

For the study of the \water\ molecule we have chosen a simple model
that is well suited for classical calculations \cite{murrell84}. The
potential energy has the form
\begin{equation}
  \label{eq:vhho}
  V_{HHO} = V_O^{(1)} \cdot f(R_1,R_2,R_3) + V_{OH}^{(2)}(R_1) 
  + V_{OH}^{(2)}(R_2) + V_{HH}^{(2)}(R_3) 
  + V_{HHO}^{(3)}(R_1,R_2,R_3)
\end{equation}
where $R_1$, $R_2$ and $R_3$ are the O--H$_1$, O--H$_2$ and
H$_1$--H$_2$ distances respectively. Distances are measured in
Angstroms and energies in eV. The single-body term in equation
\eqref{eq:vhho} is given by
\begin{equation}
  \label{eq:vo1}
   V_O^{(1)} \cdot f(R_1,R_2,R_3) = 1.958 \cdot \frac{1}{2}
   \left[
     1-\tanh\left(\frac{3\rho_3-\rho_1-\rho_2}{2}
       \medspace \alpha\right) \right]
\end{equation}
where $\alpha=1.9018$, $\rho_1=R_1-0.9572$, $\rho_2=R_2-0.9572$ and $\rho_3=
R_3-1.5139$. The two-body terms are given by the equations
\begin{equation}
  \label{eq:voh2}
  V_{OH}^{(2)}(R_i) = -D_1 (1+a_1r_i+a_2r_i^2+a_3r_i^3)
  \exp\left\{-a_1r_i\right\},
  \quad i=1,2
\end{equation}
where $r_i=R_i-0.9696$, $a_1=4.507$, $a_2=4.884$, $a_3=3.795$ and
$D_1=4.6211$,
\begin{equation}
  V_{HH}^{(2)}(R_3) = -D_2 (1+a_4r_3+a_5r_3^2+a_6r_3^3)
  \exp\left\{-a_4r_3\right\}
\end{equation}
where $r_3=R_3-0.7414$, $a_4=3.961$, $a_5=4.064$, $a_6=3.574$ and
$D_2=4.7472$.
Finally the three-body term is given by
\begin{equation}
  \label{eq:vhho3}
  V_{HHO}^{(3)}(R_1,R_2,R_3) = 0.01892 \cdot P(\rho_1,\rho_2,\rho_2)
  \prod_{i=1}^{3} (1-\tanh(\gamma_i\rho_i/2))
\end{equation}
where $\gamma_1=\gamma_2=2.6$, $\gamma_3=1.5$ and
$P(\rho_1,\rho_2,\rho_3)$ is the polynomial
\begin{multline}
  \label{eq:polynom}
  P(\rho_1,\rho_2,\rho_3) = 1 + \sum_{i=1}^{3} C_i \rho_i
  + \sum_{i=1}^{3}\sum_{j=i}^{3} C_{ij} \rho_i \rho_j \\
  + \sum_{i=1}^{3}\sum_{j=i}^{3}\sum_{k=j}^{3} C_{ijk} \rho_i \rho_j \rho_k
  + \sum_{i=1}^{3}\sum_{j=i}^{3}\sum_{k=j}^{3}\sum_{l=k}^{3} C_{ijkl}
 \rho_i \rho_j \rho_k \rho_l
\end{multline}
The coefficients $C_i$, $C_{ij}$, $C_{ijk}$ and $C_{ijkl}$ are
defined in the paper by Murrell and Carter\cite{murrell84}.

\begin{figure}\begin{center}
\includegraphics{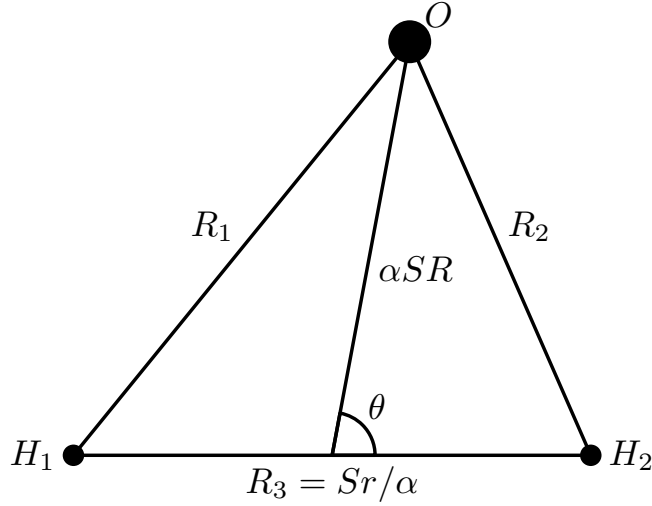}
\caption{Coordinates for the \water\ molecule.}
\label{fig:coo}
\end{center}\end{figure}

Instead of the original coordinates $R_1$, $R_2$, $R_3$ we will use
the scaled Jacobi coordinates (SJC) $SR$, $Sr$, $\theta$
(figure~\ref{fig:coo})
that are defined by the equations
\begin{equation}
  \label{eq:sjc}
  \begin{split}
    \SR &= \frac{1}{2\alpha} \sqrt{2 R_1^2 + 2 R_2^2 - R_3^2}
    \\
    \Sr &= \alpha R_3
    \\
    \cos\theta &= \frac{R_2^2-R_1^2}{2 \alpha \; SR \; R_3}
  \end{split}
\end{equation}
where $\alpha$ is given by the relation
\begin{equation}
  \label{eq:alpha}
  \alpha = \left\{ \frac{m_O+2m_H}{4m_O} \right\}^{\frac{1}{4}}
\end{equation}
The inverse relations are 
\begin{equation}
  \label{eq:sjcinv}
  \begin{split}
    R_1 &= \sqrt{\left(\frac{\Sr}{2\alpha}\right)^2 + (\alpha \; \SR)^2
      - \SR \; \Sr \cos\theta} \\
    R_2 &= \sqrt{\left(\frac{\Sr}{2\alpha}\right)^2 + (\alpha \; \SR)^2
      + \SR \; \Sr \cos\theta} \\
    R_3 &= \frac{\Sr}{\alpha} \\
  \end{split}
\end{equation}

The potential function given by equations
(\ref{eq:vhho}-\ref{eq:polynom}) can be written in scaled Jacobi
coordinates using equations \eqref{eq:sjcinv}.

The kinetic energy of the \water\ molecule has the form
\begin{equation}
  \label{eq:ken}
  T = \frac{1}{2\mu}\left[ \PSR^2 + \PSr^2
    + \left( \frac{1}{\SR^2} + \frac{1}{\Sr^2} \right) P_\theta^2
  \right]
\end{equation}
where $\PSR$, $\PSr$ and $P_\theta$ are the conjugate momenta of
$\SR$, $\Sr$ and $\theta$ respectively. The parameter $\mu$ is defined
by the relation
\begin{equation}
  \label{eq:mu}
  \mu = \sqrt{\frac{m_Om_H^2}{m_O+2m_H}} = \frac{m_H}{2\alpha^2}
\end{equation}

The Hamiltonian can now be written as the sum of the kinetic and
potential energies
\begin{equation}
  \label{eq:ham}
  H(\PSR,\PSr,P_\theta,\SR,\Sr,\theta) =
  T(\PSR,\PSr,P_\theta,\SR,\Sr) + V_{HHO}(\SR,\Sr,\theta)
\end{equation}

\section{Equipotential surfaces}

If we consider initial conditions with
\begin{equation}
  \label{eq:symman}
  \theta=\pi/2 \text{ and } P_\theta=0
\end{equation}
we can easily see that $\dot \theta=0$ and $\dot P_\theta=0$, meaning that the
submanifold of the phase space determined by equations
\eqref{eq:symman} is invariant under the Hamiltonian flow and hence we
can perform a standard type reduction in order to get a two
degrees of freedom Hamiltonian system on this submanifold.

The reduced Hamiltonian $\tilde{H}$ is related to the original
Hamiltonian through the relation
\begin{equation}
  \label{eq:redham}
  \tilde{H}(\PSR,\PSr,\SR,\Sr) = H(\PSR,\PSr,0,\SR,\Sr,\pi/2) = E
\end{equation}
and its numerical value is the energy $E$ of the molecule. This
Hamiltonian describes the symmetric water molecule, where the
distances O--H$_1$ and O--H$_2$ are equal.

\begin{figure}\begin{center}
\includegraphics{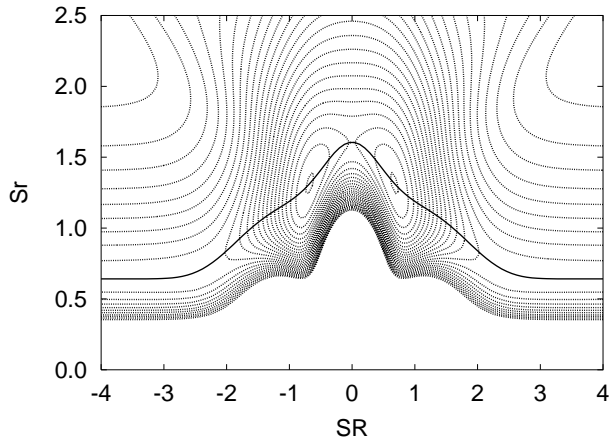}
\caption{Equipotential surfaces for the symmetric \water\
  molecule. The solid line is the Poincar{\'e} surface of section.}
\label{fig:isos}
\end{center}\end{figure}

The equipotential surfaces for the symmetric molecule are shown in
figure \ref{fig:isos}.  The potential has two minima at
$\SR=\pm0.676065$, $\Sr=1.31082$ where it takes the value
$E_0=-10.047$. Between the two minima there is a saddle point at
$\SR=0$, $\Sr=1.6057$ where the potential takes the value
$E_1=-8.993$. For $E<E_1$ orbits are confined inside the potential
well around a minimum of the potential. For $E>E_1$ the two wells join
and orbits can pass from one well to the other.

For values of energy greater than $E_2=-2.7892$ orbits can escape to
infinity following the two horizontal channels at the left and the
right of figure \ref{fig:isos}.  Escape through these channels
corresponds to a configuration of the molecule where the oxygen atom
has become unbounded and the two hydrogen atoms form an H$_2$
molecule. Although for $E>E_2$ most orbits escape there are still
bounded orbits as one can see in figure \ref{fig:psp-large}(c) below.

\section{A Poincar{\'e} surface of section}

A Poincar\'e surface of section (PSS) should intersect almost all the
orbits \cite{birkhoff27}. Such intersections occur naturally if the
potential has a plane of symmetry. However this does not happen in the
present case. On the other hand many orbits have a tendency to pass
through a local minimum of the potential (although this is not always
the case).  Thus in order to find a suitable Poincar{\'e} surface of
section we calculated numerically for each value of $\SR$ the
corresponding value of $\Sr$ where the potential function has a
minimum. Then we fitted the numerically determined function
$\Sr=f_{\text{num}}(\SR)$ by the analytic form
\begin{equation}
  \label{eq:fitsur}
  \Sr = f(SR) = \alpha \bar{f}(\alpha\;\SR)
\end{equation}
where
\begin{equation}
  \label{eq:fitfun}
  \bar{f}(x) = 0.740350
  + 0.482392 \;\exp(-3.34209\;x^2) 
  + 0.631258 \;\exp(-0.112678\;x^4)
\end{equation}
and $\alpha$ is given by equation \eqref{eq:alpha}.

This curve is given in figure \ref{fig:isos}. We checked that the
difference between the numerically determined curve
$\Sr=f_{\text{num}}(\SR)$ and the fitted analytic form is very
small. For our computations we used the fitted form, given by
equations \eqref{eq:fitsur} and \eqref{eq:fitfun}.

\begin{figure}\begin{center}
\includegraphics{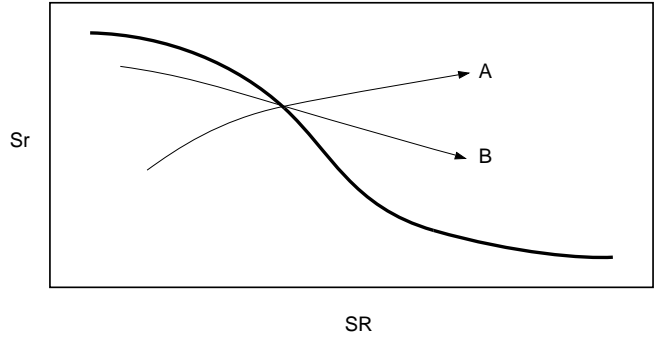}
\caption{The thick line gives the PSS. Orbits $A$ and $B$ cross the
  line at the same point with the same $\PSR$ but opposite $\PSr$.}
\label{fig:cex}
\end{center}\end{figure}

Contrary to the textbook approach we have not chosen $\SR$ and $\PSR$
as the two coordinates on the Poincar{\'e} surface of section
(PSS). If we had made such a choice we would have the following
problem. Consider the orbits of figure \ref{fig:cex}. Both orbits
cross the PSS at the same point with coordinates $\SR$, $\Sr$ with the
same momentum $\PSR$ and they both have the same energy $E$.  Then we
solve equation \eqref{eq:redham} in order to determine $\PSr$, and we
get two solutions $\pm\PSr$. From figure \ref{fig:cex} we can see that
both solutions are valid, but they correspond to different
orbits. This in turn means that on such a PSS different orbits would
overlap and this is not permissible.

\begin{figure}\begin{center}
\includegraphics{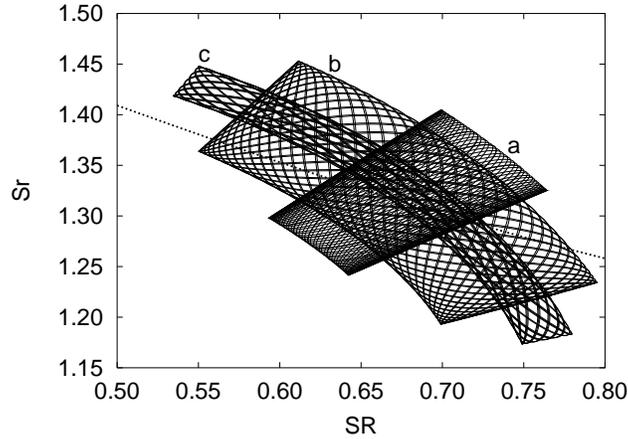}
\caption{The orbits a and c belong in the islands of 1a and 1c
  respectively. The orbit b lies between these two islands. The thick
  line represents the PSS.}
\label{fig:broken}
\end{center}\end{figure}

In order to solve this problem we have chosen as coordinates on the
PSS, $\SR$ and the component $P_t$ of the momentum vector $P=(\PSR,
\PSr)$ tangential to the curve \eqref{eq:fitsur}. Thus the Poincar\'e
surface of section is determined by the coordinate $\SR$ along the
curve of figure \ref{fig:isos} and the component $P_t$ of the
momentum. Some typical orbits on the plane $(\SR,\Sr)$ are shown in
figure \ref{fig:broken}.

A variable that is canonically conjugate to $P_t$ is
\begin{equation}
\label{qt} 
  Q_t = \int_{0}^{\SR}\sqrt{1+f'(x)} \dd x
\end{equation}
In fact the symplectic form in the variables $Q_t$, $P_t$ is
\begin{equation}
\label{omqt}
  \Omega = (\dd Q_t)\wedge (\dd P_t)
\end{equation}
and it can be shown that this is preserved.
The form \eqref{omqt} in the variables $\SR$, $P_t$ is written
\begin{equation}
\label{omsr}
  \Omega=(1+f'(\SR))^{1/2}(\dd\SR)\wedge (\dd P_t)
\end{equation}
Thus the two-form $\dd\SR\wedge \dd P_t$ is not preserved in
general. In our study we use the variable $\SR$ instead of $Q_t$
because it is not practical to compute $Q_t$ numerically.

Although the coordinates $\SR$ and $P_t$ are not canonically
conjugate, this does not cause any problems in the study of the
system. In fact the integration is performed in the canonical
coordinates $\SR$, $\Sr$, $\PSR$, $\PSr$ and the coordinates $\SR$ and
$P_t$ are used only on the PSS. The result is that the phase plots on
the PSS are only slightly distorted relative to the phase plots that
we would get if we had chosen canonically conjugate coordinates on the
PSS.

\section{Phase plots on the Poincar\'e surface of section}

\begin{figure}\begin{center}
\includegraphics[scale=1.0]{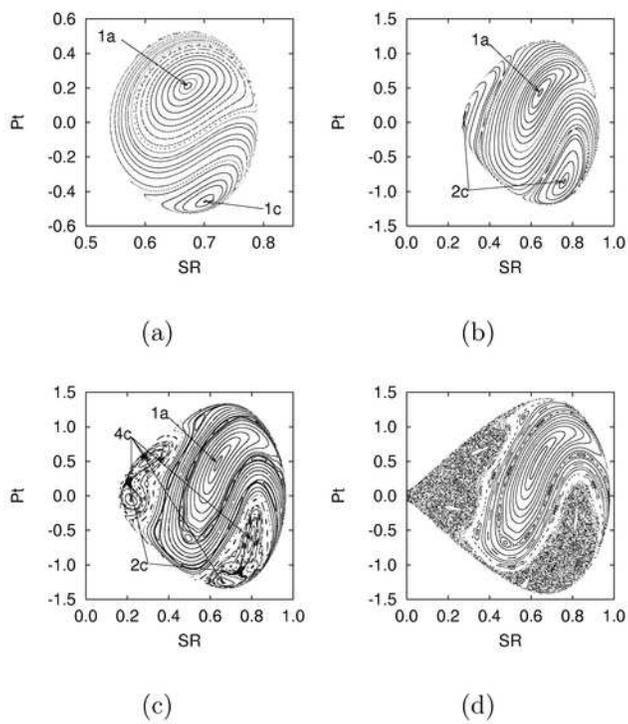}
\caption{Phase plots for small values of the energy.
(a) $E=-9.9$ (b) $E=-9.3$ (c) $E=-9.1$ (d) $E=-8.99$.}
\label{fig:psp-small}
\end{center}\end{figure}

\begin{figure}\begin{center}
\includegraphics{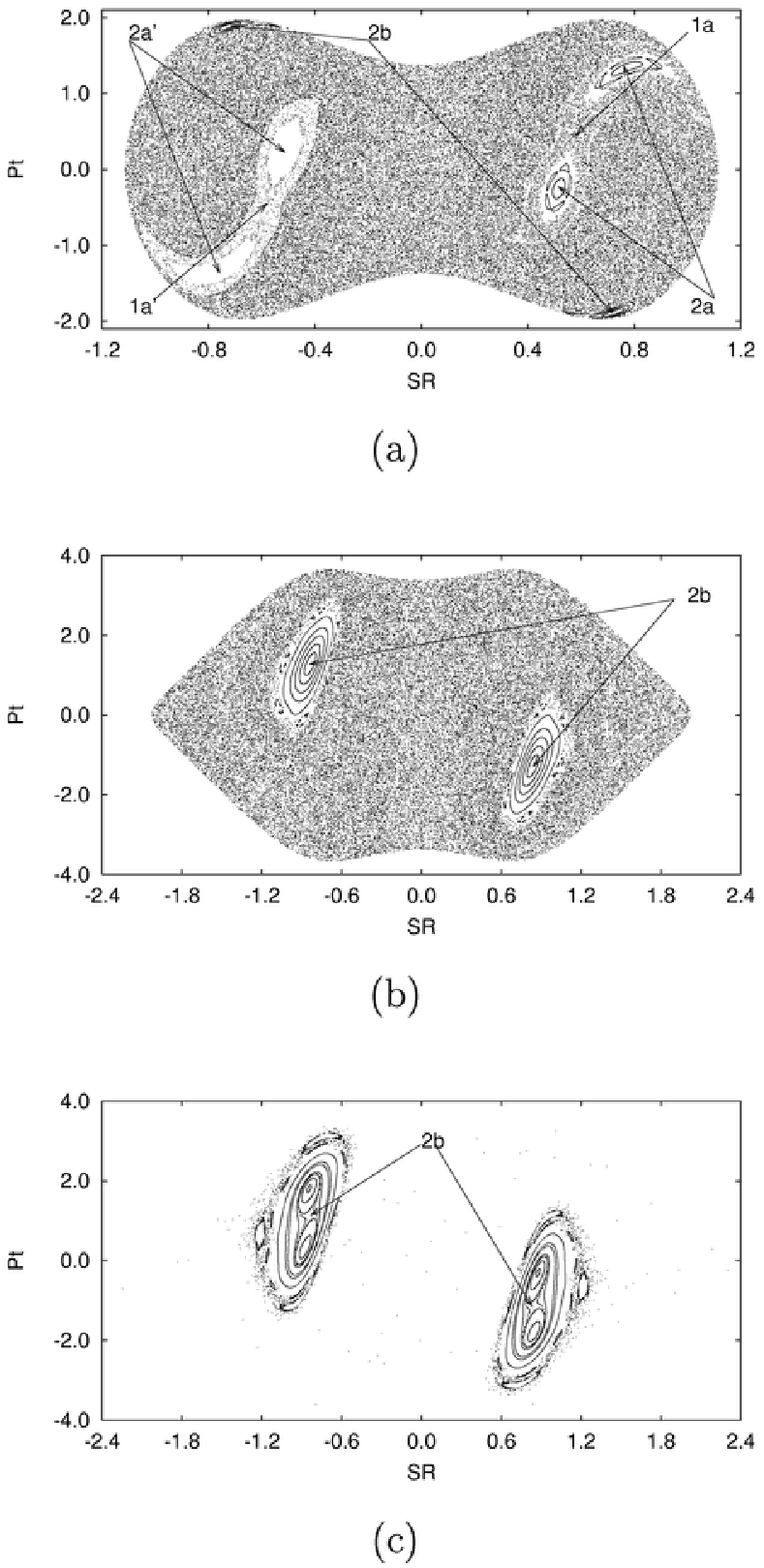}
\caption{Phase plots for energies:
(a) $E=-8.0$ (b) $E=-3.0$ (c) $E=-2.5$.}
\label{fig:psp-large}
\end{center}\end{figure}

In figures \ref{fig:psp-small}-\ref{fig:psp-large} we give the
distribution of the orbits on the Poincar\'e surface of section for
various values of the energy $E$.

The energy of figure \ref{fig:psp-small}(a) is close to the minimum of
the potential. In this plot practically all the orbits are ordered,
defining closed invariant curves.  Many invariant curves form two sets
around two stable periodic orbits.  The upper one is the orbit 1a and
the lower one the orbit 1c (see next section).  However there are also
invariant curves between these two sets, that start on the left or
upper side of the boundary, and terminate on the right side of the
boundary. In figure \ref{fig:broken} we can see why this is
happening. In this figure we see three orbits for this value of the
energy in configuration space. The orbits labeled as `a' and `c'
belong to the islands of the periodic orbits 1a and 1c
respectively. We see that the PSS crosses the outline of these orbits
at two non-adjacent sides. When this is happening we get closed
invariant curves on the PSS. But for the orbit labeled as `b', that
lies between the two islands of figure \ref{fig:psp-small}(a),
the PSS crosses the outline of this orbit at two adjacent sides
(figure \ref{fig:broken}), and on the PSS we get an open invariant
curve.

Figures \ref{fig:psp-small}(b), \ref{fig:psp-small}(c) and
\ref{fig:psp-small}(d) refer to larger energies and are given in a
larger scale than figure \ref{fig:psp-small}(a).  The orbit 1c is
continued as the double period orbit 2c in figure
\ref{fig:psp-small}(b), the second point being near the left limit of
the figure. The islands around both points 2c are now larger. In this
case also the chaotic orbits are insignificant.

In figure \ref{fig:psp-small}(c) the orbit 2c is unstable (and is
represented by two points at the centers of two dark regions) and it
has produced by bifurcation four stable points that correspond to a
stable orbit 4c of period 4, surrounded by islands of stability
above and below each point of the orbit 2c. In this case there are
some small chaotic zones around each unstable orbit.

The chaotic zones have grown considerably in figure
\ref{fig:psp-small}(d). In this case there are still 4 islands of
stability 4c, as in figure \ref{fig:psp-small}(c), but very
small. All the rest of the area previously occupied by bifurcations of
the orbit 2c and their corresponding islands are now chaotic. This
case is just above the limiting value of the energy $E_1=-8.993$ at
which the two potential wells join into one. Thus there is one
symmetric figure for $\SR<0$ that joins the figure
\ref{fig:psp-small}(d) at the small throat near the point $(0,0)$.
The point $(0,0)$ represents the orbit 1b, which is unstable and
produces a large chaotic region around it.

In figure \ref{fig:psp-large}(a) we give the phase plot for a larger value
of the energy ($E=-8$) when the two potential wells are well connected
at $\SR=0$ (see the equipotentials of
figure~\ref{fig:isos}). Figure~\ref{fig:psp-large}(a) is symmetric with
respect to the axis $\SR=0$. The apparent asymmetries in the islands
of this figure are due to the use of non-symmetric initial conditions
on both sides of the axis $\SR=0$. In this case the orbit 1a is
unstable and has generated by bifurcation the orbit 2a, surrounded by
two islands of stability on the right half of figure
\ref{fig:psp-large}(a). A symmetric orbit 1a' and its bifurcations 2a'
appear on the left side of the figure. The orbits 1a, 1a' have
generated two chaotic domains, that surround the islands 2a
(respectively 2a'). At the lower left and upper right sides of this
figure there are two elongated islands that belong to the family 2b.
The center of the figure $(0,0)$ represents again the unstable orbit
1b, that generates a large chaotic domain.

Most of figure~\ref{fig:psp-large}(a) is covered by one chaotic orbit for
which we have calculated $6\cdot10^4$ points.  We see that the points
of the chaotic orbit are not distributed uniformly on the PSS. This is
a transient phenomenon and is due to stickiness of the orbits near
cantori surrounding the islands 2a, or the islands 2a'.  These cantori
 form partial barriers. Orbits
need a long time in order to cross the cantori and for considerable
times the regions inside and outside the cantori seem separated. This
causes the apparent non uniformity of the distribution of points on
the PSS. But after a larger number of iterations the distribution of
the points becomes uniform.

The phase plots for larger values of $E$ have some similarity with
figure~\ref{fig:psp-large}(a) but there are also differences. For example in
figure~\ref{fig:psp-large}(b) we see that there are two symmetric islands,
one on each side of the axis $\SR=0$. These islands belong to the
irregular family 2b, which is symmetric with respect to the
point $(0,0)$ (see next section), and not to two different families
like the families 1a and 1a' of figure \ref{fig:psp-large}(a). The chaotic
domain in this case is again very large.

For $E$ above the energy $E=E_2$ orbits can escape. In
figure~\ref{fig:psp-large}(c) we can see the islands around an orbit that
belongs in the family 2b. The orbits that are outside these islands
escape and this is why the area between the islands is empty.

\begin{figure}\begin{center}
\includegraphics{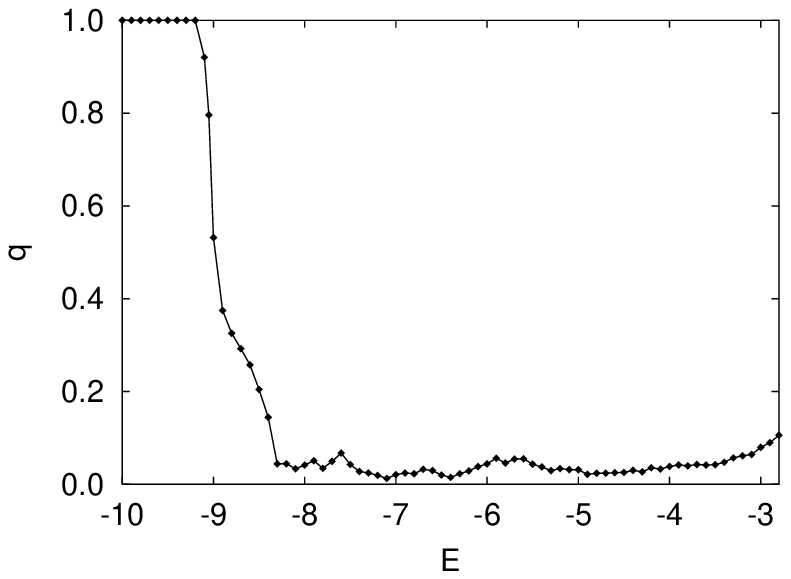}
\caption{Proportion of regular orbits, as a function of the energy $E$}
\label{fig:emb}
\end{center}\end{figure}

The percentage of the area of the PSS covered by organized orbits as a
function of the energy is presented in figure~\ref{fig:emb}. We see
that for values of energy close to $E_1=-10.047$ (the minimum value of
the potential) almost all orbits are organized. An abrupt change
happens at values of energy around $E=-9$. For $E>-9$ most of the PSS
is filled by chaotic orbits. There are islands of stability for even
larger values of the energy, and we notice that for values of the
energy close to $E=-3$ the percentage of the organized orbits
increases, although it remains below $0.2$ (or $20\%$). This increase
is due to the increase of the size of the island around the family
2b. The islands continue to increase in size even when $E$ becomes
larger than the escape energy $E_2=-2.79$, but for still larger $E$
they become smaller.

\section{Periodic orbits}

\begin{figure}\begin{center}
\includegraphics{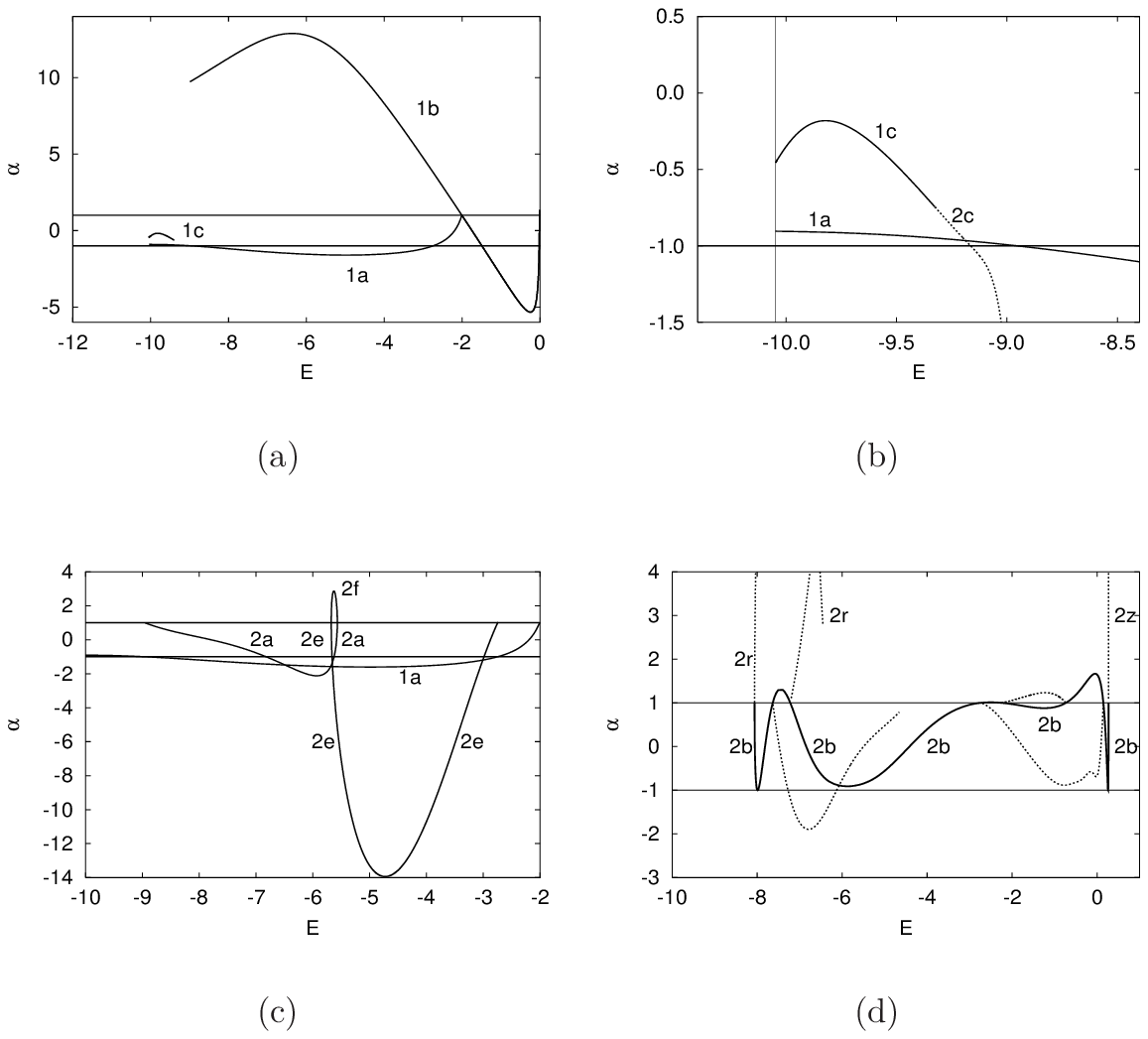}
\caption{Stability diagrams: (a) Families 1a, 1b, 1c (b) Families 1a,
  1c, 2c (continuation of 1c) (c) Family 1a and its period 2
  bifurcations (d) Families 2b, 2r, 2z.}
\label{fig:stabdiag}
\end{center}\end{figure}

We have computed the most important families of periodic orbits for
the symmetric water molecule. The stability diagram of the period-1
families as a function of the energy of the molecule can be seen in
figure \ref{fig:stabdiag}(a). Such a diagram gives the H\'enon stability
parameter \cite{henon65} $\alpha$ of each family as a function of $E$.
For a period-$k$ orbit of an area-preserving map $\sigma$, the H\'enon
stability parameter is defined by the relation
\begin{equation}
  \label{eq:hsidef}
  \alpha = \frac{1}{2}\text{Trace}(D[\sigma^k](x_0))
\end{equation}
where $x_0$ is a point of the periodic orbit and $D[\sigma^k](x_0)$ is
the Jacobian matrix of $\sigma\circ \cdots \circ \sigma$ ($k$~times)
evaluated at $x_0$.  When the stability parameter of a periodic orbit
lies in the interval from $-1$ to $1$ the orbit is stable, otherwise
it is unstable.

Since the Poincar\'e map is not symplectic, as we have explained in
section~4, the determinant of the Jacobian matrix $D\sigma(x)$ is not
equal to $1$ for arbitrary points $x$ on the PSS. However for a
periodic orbit with period $k$ the determinant of $D[\sigma^k]$ is
exactly equal to $1$. In fact, using equation \eqref{omsr} it can be
proven that the determinant of the Jacobian matrix of $\sigma^k$ at a
point $x_0=(\SR_0,P_{t0})$ is given by the relation
\begin{equation}
  \det D[\sigma^k](x_0) = \sqrt{\frac{1+f'(SR_0)}{1+f'(SR_k)}}
\end{equation}
where $x_k=\sigma^k(x_0)=(\SR_k,P_{tk})$ is the $k$-th image of $x_0$.
If an orbit is periodic with period $k$, then $\SR_0=\SR_k$ hence
$\det D[\sigma^k](x_0)=1$.

\begin{figure}\begin{center}
\includegraphics{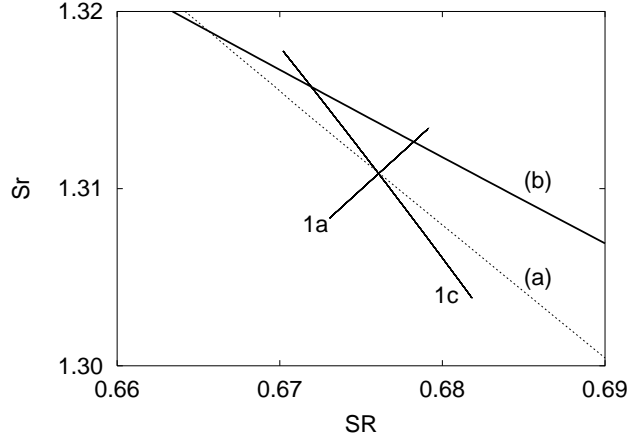}
\caption{Periodic orbits 1a and 1c for $E=-10.04664$ in configuration
  space. The line (a) represents the numerically determined minima
  of the potential and the line (b) represents the selected
  PSS.}
\label{fig:p1ac100}
\end{center}\end{figure}

For values of energy close to the minimum value of the potential
($E=E_1=-10.047$) there exist two period-1 families. We call these
families 1a and 1c. They correspond to the two normal modes of the
water molecule. In figure~\ref{fig:p1ac100} we can see the
corresponding periodic orbits at $E=-10.04664$ in configuration space.

\begin{figure}\begin{center}
\includegraphics{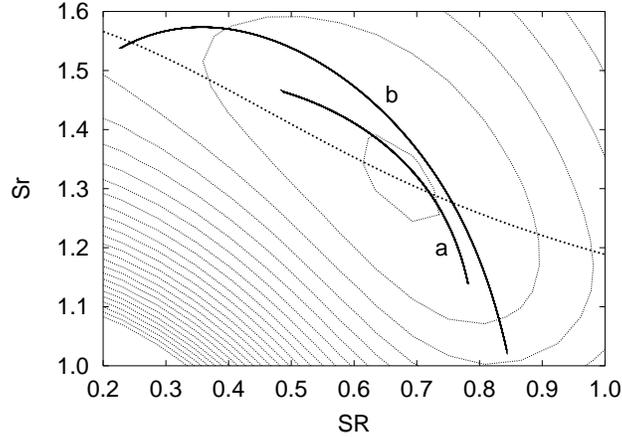}
\caption{The transition 1c$\rightarrow$2c. The orbit `a' ($E=-9.8$)
belongs to the  family 1c. The orbit `b' ($E=-9.2$) belongs to the
family 2c.}
\label{fig:p12cx}
\end{center}\end{figure}

The family 1c has a peculiar behaviour that is the result of our
choice of the PSS. This family is born at $E=-10.047$, but beyond the
energy $E=-9.331$ it appears as a period-2 family, that we call 2c
(figure \ref{fig:stabdiag}(b)). We note that no bifurcation appears
here. In order to understand what is happening, we should see figure
\ref{fig:p12cx}. In this figure we see one orbit of the family 1c that
intersects the PSS only at one point, as expected for small
energies. An orbit of the same family for a larger value of the energy
intersects the PSS at two points and thus it appears as a period-2
orbit. We will call the change of the apparent period of a family that
is not related to a bifurcation, a \emph{transition}.  We note here
that there would be a transition even if we had chosen, as the PSS, the
numerically determined curve of the minima of the potential, instead
of the fitted function.  Family 2c becomes unstable at $E=-9.164$ and
generates by bifurcation two stable period 4 families.

For values of the energy between $E=E_1=-10.047$ and $E=-7.2$ the
largest islands on the PSS are around the family 1a and the family 2a
that bifurcates from 1a (figures \ref{fig:psp-small} and
\ref{fig:psp-large}(a)). The family 1a is created at the minimum energy of
the system $E=E_1=-10.047$ as a stable family and becomes unstable at
$E=-8.95$ (figure~\ref{fig:stabdiag}(a)). At this value of $E$ it generates
by bifurcation the period-2 family 2a
(figure~\ref{fig:stabdiag}(c)). Notice that when the stability parameter
$\alpha$ of a period-1 orbit is equal to $-1$ the same orbit described
twice has a stability parameter $+1$. The family 2a starts at
$E=-8.95$ with stability parameter $\alpha=1$, and exists for larger
values of $E$. The family 1a remains unstable until $E=-2.47$, where
the period-2 family 2e is generated by bifurcation from the family 1a,
existing for smaller values of $E$ (figure~\ref{fig:stabdiag}(c)).

\begin{figure}\begin{center}
\includegraphics{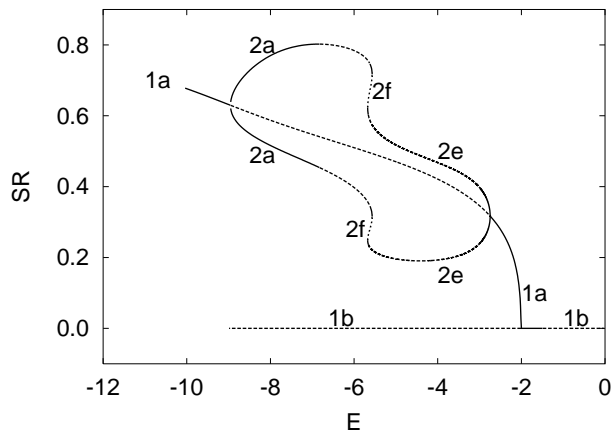}
\caption{Bifurcation diagram of family 1a}
\label{fig:bdp1asre}
\end{center}\end{figure}

The family 2a for $E=-6.82$ becomes unstable by crossing the line
$\alpha=-1$ and becomes stable again for $E=-5.62$. The stability
parameter of the family 2a takes the value $\alpha=1$ at
$E=-5.568$. At this energy an unstable period-2 family, that we call
2f is created.  The family 2e becomes unstable (going from larger to
smaller values of the energy) for $E=-2.987$ and becomes stable again
for $E=-5.661$.  The stability parameter of the family 2e takes the
value $\alpha=1$ at $E=-5.675$.  At this energy the unstable family
2f, that was created at a larger energy from 2a, joins 2e
(figure~\ref{fig:stabdiag}(c)). The bifurcation diagram of the families
1a, 2a, 2e, 2f and 1b is shown in figure~\ref{fig:bdp1asre}. This
figure gives the value of $\SR$ for each energy. We see that the
family 2f exists only in a small interval of values of $E$ and joins
the family 1a at its maximum $E$ and the family 2e at its minimum
$E$. This explains why the families 2a, 2e and 2f form a loop in
figure~\ref{fig:stabdiag}(c).

Between $E=-2.47$ and $E=-2.00$ the family 1a remains stable. At
$E=-2.00$ the family 1a joins the family 1b, and for larger values of
energy the family 1a does not exist. We may thus say that the family
1a is a bifurcation of the family 1b. The family 1b is represented by
the point $(0,0)$ on the surface of section. It is created at the
energy $E_2=-8.993$ (figure \ref{fig:stabdiag}(a)), at the saddle point of
the potential (figure \ref{fig:isos}), with coordinates $\SR=0$,
$\Sr=1.6057$. For energies somewhat larger than $E_2$ the family 1b is
unstable ($\alpha=9.73$). As $E$ increases, the stability parameter of
the family 1b becomes larger, but later (for larger $E$) it becomes
smaller. At $E=-2.00$ this family becomes stable. It becomes again
unstable at $E=-1.50$ with stability parameter $\alpha=-1$. For larger
$E$ the family 1b is unstable, but at $E=-0.0205$ it becomes stable
again until $E=-0.0110$, where it becomes unstable, by crossing the
line $\alpha=1$ and thus generating another period-1 family. As the
energy approaches the energy $E=0$ the stability parameter of the
family 1b tends to infinity, and the return time (period) of the
periodic orbit also tends to infinity.

\begin{figure}\begin{center}
\includegraphics{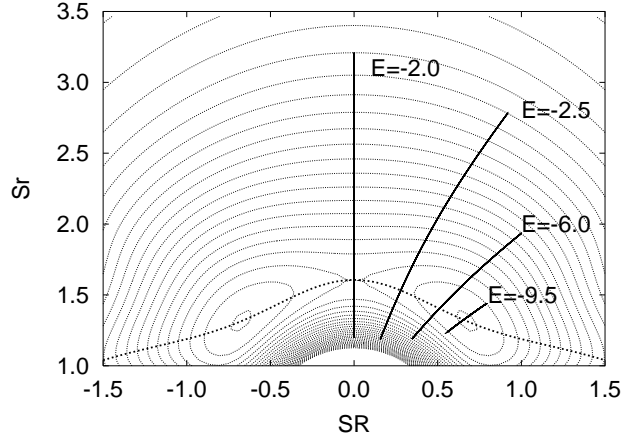}
\caption{Periodic orbits of the families 1b (axis $\SR=0$) and 1a.
  The dotted line represents the PSS.}
\label{fig:p1acs}
\end{center}\end{figure}

In figure~\ref{fig:p1acs} we can see that periodic orbits of the
family 1b appear as straight lines in configuration space with $\SR=0$
and varying $\Sr$. In physical space this corresponds to a colinear
configuration of the molecule with the oxygen atom in the middle and
the two hydrogen atoms oscillating symmetrically at each side of the
oxygen atom. For large intervals of the energy this configuration is
unstable. In figure~\ref{fig:p1acs} we can also see how the periodic
orbits of the family 1a appear in configuration space for different
values of the energy. We note that there is also a family 1a' that is
symmetric to the family 1a with respect to the axis $\SR=0$. As the
energy increases the families 1a and 1a' approach the
family 1b, and join 1b for $E=-2.00$.

At energy $E=-8.06$ two irregular period-2 families are created
(figure \ref{fig:stabdiag}(d)).  Irregular families are created in pairs
(one stable and one unstable) by a tangent bifurcation. This means
that irregular orbits are not created by the bifurcation of an
existing periodic orbit, but from each other. The stable family is
called 2b and the unstable one is called 2r. The stability parameter
of the family 2r grows from $\alpha=1$ to $\alpha=17.12$ as the energy
increases and then decreases, until it reaches the value $\alpha=2.77$
at $E=-6.454$. For this energy the family 2r appears as a period-4
family and we have a transition, as in the case of the transition
1c$\rightarrow$2c.

The stable family 2b plays an important role in the structure of the
phase space. The stability diagram of the family 2b and the period-2
families that bifurcate from 2b can be seen in
figure~\ref{fig:stabdiag}(d). The diagram is very complex, much more than
the stability diagram of the family 1a. Here we consider only the
family 2b and its equal period bifurcations.  The stability parameter
of the family 2b crosses (or just reaches) the horizontal line
$\alpha=1$ eight times. Each time this happens a family of period-2
orbits bifurcates from the family 2b. For $E=0.27$ the
family 2b joins a new unstable family 2z and they both disappear
there, i.e. they do not exist for larger $E$.

\begin{figure}\begin{center}
\includegraphics{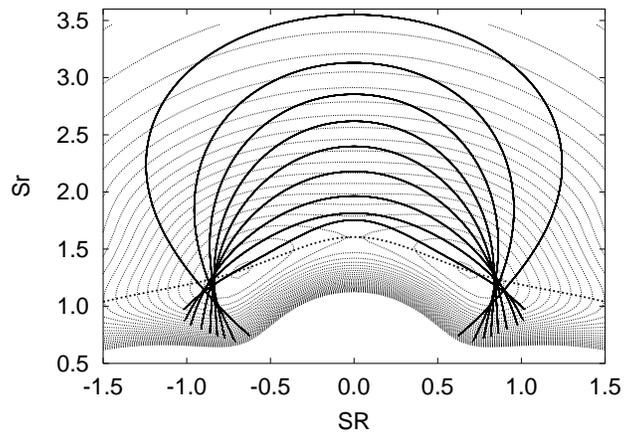}
\caption{Orbits of the family 2b in configuration space. The dotted
  line represents the PSS.}
\label{fig:p2bcs}
\end{center}\end{figure}

The families 2b, 2r and 2z are symmetric with respect to the point
$(0,0)$. The families that bifurcate from 2b are not symmetric but
they come in symmetric pairs. In figure~\ref{fig:stabdiag}(d) some families
that have bifurcated from the family 2b suddenly terminate because
they are going through transitions. Some orbits of the family 2b can
be seen in figure~\ref{fig:p2bcs}. In physical space these orbits
correspond to oscillations of the oxygen atom between the two
symmetric minima of the potential.

The island around the orbits of the family 2b is the largest island
for values of energy close to $E=-3$. Generally, from $E=-8$ to $E=-3$ 
the island around 2b occupies a significant measure of the PSS.

\begin{figure}\begin{center}
\includegraphics{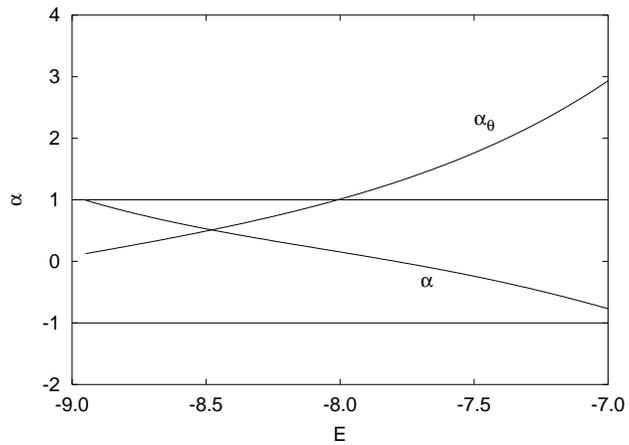}
\caption{Stability diagrams of the family 2a for deviations on the
  symmetry plane ($\alpha$) and perpendicular to it
  ($\alpha_\theta$).}
\label{fig:last}
\end{center}\end{figure}

In the present paper we studied the stability (and instability) of the
various types of orbits on the plane of symmetry $\theta=\pi/2$ and
$P_\theta=0$. However orbits that are stable on this plane may be
unstable in a direction perpendicular to this plane. An example is
shown in figure \ref{fig:last}, where we have calculated two stability
parameters for the family 2a. One is the usual stability parameter
$\alpha$, as in figure \ref{fig:stabdiag}(c), and the other is the
stability parameter $\alpha_\theta$ for deviations out of the symmetry
plane. We see that while the family 2a is stable with respect to
deviation on the symmetry plane it becomes unstable with respect to
perpendicular deviations for $E=-8$. At the transition to instability
there is a bifurcation of a new period 2 family, which extends out of
the symmetry plane and is initially stable.

The study of the stability of orbits in a direction out of the
symmetry plane is important in many cases. Such studies in galactic
dynamics have been made since a long time\cite{contopoulos85}. A
detailed study of the stability and the bifurcations of periodic
orbits with respect to the third dimension in the \water\ molecule
will be given in a future paper.

\section{Summary and conclusions}

We have made a systematic study of the classical orbits for the
symmetric configuration of a realistic model of the \water\
molecule.

In order to study the distribution of the ordered and the chaotic
orbits for this system, we had to choose an appropriate Poincar\'e
surface of section (PSS). The usual choice for the PSS is a plane
surface in phase space, but for the particular system under study here,
this would not be convenient, because for any choice of a plane
PSS there would be important periodic orbits that would not cross it.

The PSS we chose is a surface passing approximately through the minima
of the potential. Most non-escaping orbits cross this PSS. Choosing
appropriate coordinates on this PSS, we were able to reduce the study
of the dynamics of the symmetric \water\ molecule to the study of a
two-dimensional map. With this particular choice of the PSS we ensure
that the Poincar\'e map describes the complete dynamics of the
molecule.  A minor problem with this particular choice for the PSS is
that some periodic families appear to change period without a
bifurcation, due to an extra intersection of a periodic orbit with the
PSS.

We found the distribution of the orbits on this PSS, distinguishing
between ordered and chaotic orbits. The ordered orbits are represented
by isolated points if they are periodic and by invariant curves if
they are quasi-periodic. Most orbits are ordered for small
energies. But as the energy increases beyond a critical value the
proportion of ordered orbits decreases abruptly. When the energy
increases beyond the escape energy ($E_2=-2.7892$), most orbits escape
to infinity, but there are still orbits trapped around stable periodic
orbits. This remains true even when the energy increases above
$E_3=0$, but for positive energies the regions of ordered motion are
very small, and practically insignificant.

We studied the main periodic orbits of period 1 and 2 for negative
values of the energy. We constructed stability diagrams and
bifurcation diagrams for the various families of periodic
orbits. These diagrams are essential in understanding the role of
various stable orbits in trapping non-periodic orbits around them. In
fact for different energies the trapping takes place around different
orbits and we can separate the interval between the critical energy
and the escape energy into two regions. In the first region (lower
energies) the ordered orbits are trapped mainly around the families 2a
and 2a'. In the second region, close to the escape energy the ordered
orbits are trapped mainly around the irregular family 2b.

\section*{Acknowledgments}

This research was supported by the Research Committee of the Academy
of Athens (Grant No. 200/419).  We thank Prof. S. Farantos for his
suggestion to use the particular model studied here. We also thank
Dr. C. Skokos for useful discussions. KE has been supported in part by
the Greek Foundation of State Scholarships (IKY). We thank also the
(anonymous) referees for their remarks and for providing useful
references.

\end{document}